\documentclass[twocolumn,showpacs,aps,prl,superscriptaddress,letterpaper]{revtex4}
\usepackage{graphicx}
\usepackage{dcolumn}
\usepackage{amsmath}
\usepackage{epsfig}
\long\def\inst#1{\par\nobreak\kern 4pt\nobreak
    {\itshape #1}\par\vskip 10pt plus 3pt minus 3pt}
\RequirePackage{xspace}
\usepackage{relsize}

\def\babar{\mbox{\slshape B\kern-0.1em{\smaller A}\kern-0.1em
    B\kern-0.1em{\smaller A\kern-0.2em R}}}
\def\Dbar    {\kern 0.18em\overline{\kern -0.18em D}{}\xspace}
\def\Bbar    {\kern 0.18em\overline{\kern -0.18em B}{}\xspace}

\def\BB      {\ensuremath{B\Bbar}\xspace} 
\def\Bz      {\ensuremath{B^0}\xspace}
\def\Bzb     {\ensuremath{\Bbar^0}\xspace}
\def\BzBzb   {\ensuremath{\Bz {\kern -0.16em \Bzb}}\xspace}
\def\Bu      {\ensuremath{B^+}\xspace}
\def\Bub     {\ensuremath{B^-}\xspace}

\def\BpBm    {\ensuremath{\Bu {\kern -0.16em \Bub}}\xspace}

\newcommand{\optbar}[1]{\shortstack{{\tiny (\rule[.4ex]{1em}{.1mm})}
  \\ [-.7ex] $#1$}}
\def\BorBbar    {\kern 0.18em\optbar{\kern -0.18em B}{}\xspace}
\def\DorDbar    {\kern 0.18em\optbar{\kern -0.18em D}{}\xspace}
\def\KorKbar    {\kern 0.18em\optbar{\kern -0.18em K}{}\xspace}

\def\pep2{PEP-II}
\mathchardef\Upsilon="7107
\def\Y#1S{\ensuremath{\Upsilon{(#1S)}}\xspace}

\def\FourS {\Y4S}

\def\invfb   {\ensuremath{\mbox{\,fb}^{-1}}\xspace}
\newcommand{\mev}{\ensuremath{\mathrm{\,Me\kern -0.1em V}}\xspace}

\def\mes        {\mbox{$m_{\rm ES}$}\xspace}
\def\DeltaE     {\mbox{$\Delta E$}\xspace}

\newcommand{\jprlBase}       {Phys.\ Rev.\ Lett.\xspace}
\newcommand{\jprl}      [1]  {\jprlBase\ {\bf #1}}
\newcommand{\jplBase}        {Phys.\ Lett.\xspace}

\newcommand{\jprBase}        {Phys.\ Rev.\xspace}
\newcommand{\jprd}      [1]  {\jprBase\ D~{\bf #1}}

\newcommand{\BABARPubYear}    {06}
\newcommand{\BABARPubNumber}  {050}

\newcommand{\SLACPubNumber}   {11997}

\newcommand{\rhoK}
{$B\rightarrow \rho K^*$}

\newcommand{\rhopKz}
{$B^+\rightarrow \rho ^+ {K^*}^0$}
\newcommand{\rhozKz}
{$B^0\rightarrow \rho ^0 {K^*}^0$}
\newcommand{\rhopKm}
{$B^0\rightarrow \rho ^- {K^*}^+$}
\newcommand{\fKp}
{$B^+\rightarrow f_0(980) {K^*}^+$}
\newcommand{\fKz}
{$B^0\rightarrow f_0(980) {K^*}^0$}
\newcommand{\rhozKpshort}
{$\rho ^0 {K^*}^+$}
\newcommand{\rhozKpshorta}
{$\rho ^0 {K^*}^+_{K^+\pi^0}$}
\newcommand{\rhozKpshortb}
{$\rho ^0 {K^*}^+_{K^0_S\pi^+}$}
\newcommand{\rhopKzshort}
{$\rho ^+ {K^*}^0$}
\newcommand{\rhozKzshort}
{$\rho ^0 {K^*}^0$}
\newcommand{\rhopKmshort}
{$\rho ^- {K^*}^+$}
\newcommand{\fKpshort}
{$f_0(980) {K^*}^+$}
\newcommand{\fKpshorta}
{$f_0(980) {K^*}^+_{K^+\pi^0}$}
\newcommand{\fKpshortb}
{$f_0(980) {K^*}^+_{K^0_S\pi^+}$}
\newcommand{\fKzshort}
{$f_0(980) {K^*}^0$}

\begin{document}

\begin{flushleft}
\babar-PUB-\BABARPubYear/\BABARPubNumber\\
SLAC-PUB-\SLACPubNumber\\
[10mm]
\end{flushleft}

\title{
\large \bfseries \boldmath
Measurements of branching fractions, polarizations, and direct $CP$-violation asymmetries in 
$B \rightarrow \rho K^*$ and $B \rightarrow f_0(980)K^*$ decays
}


\date{\today}
%
\author{B.~Aubert}
\author{R.~Barate}
\author{M.~Bona}
\author{D.~Boutigny}
\author{F.~Couderc}
\author{Y.~Karyotakis}
\author{J.~P.~Lees}
\author{V.~Poireau}
\author{V.~Tisserand}
\author{A.~Zghiche}
\affiliation{Laboratoire de Physique des Particules, F-74941 Annecy-le-Vieux, France }
% ...
\collaboration{The \babar\ Collaboration}
\noaffiliation

\begin{abstract}

We report searches for $B$-meson decays to the charmless final states
$\rho K^*$ and $f_0(980) K^*$ with a sample of 232 million \BB pairs collected with the
\babar\ detector at the PEP-II asymmetric-energy $e^+e^-$ collider at SLAC. 
We measure the following branching fractions in units of 
$10^{-6}$:
${\cal B} (B^+\rightarrow \rho^0 {K^*}^+) = 3.6 \pm 1.7 \pm 0.8 \;(< 6.1)$,
${\cal B} (B^+\rightarrow \rho^+ {K^*}^0) = 9.6 \pm 1.7 \pm 1.5 $,
${\cal B} (B^0\rightarrow \rho^- {K^*}^+) = 5.4 \pm 3.6 \pm 1.6 \;(< 12.0)$,
${\cal B} (B^0\rightarrow \rho^0 {K^*}^0) = 5.6 \pm 0.9 \pm 1.3 $,
${\cal B} (B^+\rightarrow f_0(980) {K^*}^+) = 5.2 \pm 1.2 \pm 0.5 $, and
${\cal B} (B^0\rightarrow f_0(980) {K^*}^0) = 2.6 \pm 0.6 \pm 0.9 \;(< 4.3)$.
The first error quoted is statistical, the second systematic, and the upper limits,
in parentheses, are given at the 90\% confidence level.
For the statistically significant modes we also measure the fraction of longitudinal polarization
and the charge asymmetry:
$f_L (B^+\rightarrow \rho^+ {K^*}^0) = 0.52 \pm 0.10 \pm 0.04 $,
$f_L (B^0\rightarrow \rho^0 {K^*}^0) = 0.57 \pm 0.09 \pm 0.08 $,
${\cal A}_{\rm CP} (B^+\rightarrow \rho^+ {K^*}^0) =  -0.01 \pm 0.16 \pm 0.02 $, 
${\cal A}_{\rm CP} (B^0\rightarrow \rho^0 {K^*}^0) =   0.09 \pm 0.19 \pm 0.02 $,
${\cal A}_{\rm CP} (B^+\rightarrow f_0(980){K^*}^+) = -0.34 \pm 0.21 \pm 0.03 $, and
${\cal A}_{\rm CP} (B^0\rightarrow f_0(980){K^*}^0) = -0.17 \pm 0.28 \pm 0.02 $.

\end{abstract}


\maketitle

The study of $B$-meson decays to charmless hadronic final states plays
an important role in understanding $CP$ violation.
The charmless decays \rhoK\ proceed through dominant penguin loops and 
Cabibbo-suppressed tree processes (\rhopKz\ is pure penguin) to two vector 
particles (VV).  A large longitudinal polarization 
fraction $f_L$ (of order $(1-4m_V^2/m_B^2)\sim 0.9$) is predicted
for both tree and penguin dominated VV decays~\cite{bib:prediction}.  
However, recent 
measurements of the pure penguin VV decays $B\rightarrow \phi K^*$
indicate $f_L\sim 0.5$~\cite{bib:phiKst}.
Several attempts to understand this small value of $f_L$
within or beyond the Standard Model (SM) have been made~\cite{bib:theory1}. 
Further information about $SU(3)$-related decays may provide some insight 
into this polarization puzzle.
Characterization of the four \rhoK\ modes can also be used within the SM 
framework to help constrain the angles $\alpha$ and $\gamma$ of the Unitarity 
Triangle~\cite{bib:theory2}.

We report measurements of branching fractions, longitudinal polarizations, 
and direct $CP$-violating asymmetries for the \rhoK\ decay modes. We also 
measure branching fractions and direct $CP$-violating asymmetries for the 
$B\to f_0(980) K^*$ modes that share the same final states. We present 
improved analyses of previously measured modes~\cite{bib:prevmeas}, with 
larger statistics and explicit consideration of non-resonant backgrounds. 
We measure \rhopKm , \rhozKz , \fKp , and \fKz\ for the first time.
Charge-conjugate modes are implied throughout.

This analysis is based on a data sample of 232 million \BB pairs,
corresponding to an integrated luminosity of 210 \invfb , 
collected with the \babar\ detector~\cite{bib:babar} at the SLAC PEP-II 
asymmetric-energy $e^+e^-$ collider operating at a center-of-mass 
(c.m.) energy $\sqrt{s} = 10.58$\,GeV, corresponding to the  
\FourS\-resonance mass.

The angular distribution of the $\rho K^*$ decay products, after integrating
over the angle between the decay planes of the vector mesons,
for which the acceptance is uniform, is proportional to
\begin{equation}
\frac{1}{4}
(1-f_L)\sin^2\theta_{K^*}\sin^2\theta_{\rho} + 
   f_L \cos^2\theta_{K^*}\cos^2\theta_{\rho} ,
\label{eq:helicity}
\end{equation}

\noindent where $\theta_{K^*}$ and $\theta_{\rho}$ are the helicity angles
of the $K^*$ and $\rho$, defined between the $K(\pi^+)$ momentum and the
direction opposite to the $B$ in the $K^*(\rho)$ rest frame~\cite{bib:polarization}.
We also measure the time-integrated direct $CP$-violating asymmetry
${\cal A}_{\rm CP} = (\Gamma^- - \Gamma^+)/(\Gamma^- + \Gamma^+)$, where
the superscript on the total width $\Gamma$ indicates the sign of 
the $b$-quark charge in the $B$ meson.

We fully reconstruct charged and neutral decay products including the
intermediate states
$\rho^0 {\;\rm or \;}f_0(980) \rightarrow \pi^+ \pi^-$,
$\rho^+  \rightarrow \pi^+ \pi^0$,
${K^*}^0 \rightarrow K^+\pi^-$,
${K^*}^+ \rightarrow K^+ \pi^0$,
${K^*}^+ \rightarrow K^0_S \pi^+$ (only in \rhozKpshort),
$\pi^0   \rightarrow \gamma \gamma$, and
$K^0_S   \rightarrow \pi^+\pi^-$.
We assume the $f_0(980)$ measured lineshape~\cite{bib:e791}
and a branching ratio of 100\% for $f_0(980) \rightarrow \pi^+ \pi^-$. 
Table~\ref{tab:resonances} lists the selection requirements on the
invariant mass and helicity angle of $B$-daughter resonances.

\begin{table}[htb]
\caption{Selection requirements on the invariant mass (in GeV) and helicity 
angle of $B$-daughter resonances.}
\begin{center}
\begin{tabular}{lcccc}
\hline
\hline
Mode           &   $m_{\pi\pi}$   &  $m_{K\pi}$    & $\cos\theta_{\rho}$  &  $\cos\theta_{K^*}$   \\
\hline
\rhozKpshorta  &  (0.52,1.10)    &  (0.75,1.05)  &   (-0.95,0.95)      &   (-0.5,1.0)      \\  
\rhozKpshortb  &  (0.52,1.10)    &  (0.75,1.05)  &   (-0.95,0.95)      &   (-0.9,1.0)      \\  
\rhopKzshort   &  (0.40,1.15)    &  (0.77,1.02)  &   (-0.66,0.95)      &   (-0.95,1.0)     \\  
$\rho ^- {K^*}^+_{K^+\pi^0}$ &  (0.40,1.15)    &  (0.77,1.02)  &   (-0.80,0.98)      &   (-0.80,0.98)    \\  
\rhozKzshort   &  (0.52,1.15)    &  (0.77,1.02)  &   (-0.95,0.95)      &   (-0.95,1.0)     \\  
\hline
\hline
\end{tabular}
\label{tab:resonances}
\end{center}
\end{table}

Charged tracks from the $B$-meson candidate are required to originate from the interaction
point. Looser criteria are applied to tracks forming $K^0_S$ candidates, for 
which we require $|m_{\pi^+\pi^-}-m_{K^0_S}|<12$\,MeV, a measured proper decay 
time greater than five times its uncertainty, and the cosine of the angle 
between the reconstructed flight and momentum directions to exceed 0.995.
Charged particle identification provides discrimination between kaons and 
pions, and is also used to reject electrons and protons.
We reconstruct $\pi^0$ mesons from pairs of photons, each with a minimum 
energy of 30\,MeV (\rhozKpshort ) or 50\,MeV (\rhopKzshort\ and \rhopKmshort ).
The invariant mass of $\pi^0$ candidates is required to be within
15\,MeV (\rhozKpshort ) or 25\,MeV (\rhopKzshort\ and \rhopKmshort ) of the 
nominal mass~\cite{bib:PDG}.

$B$-meson candidates are characterized kinematically by the energy difference
 $\Delta E = E^*_B - \sqrt{s}/2$ and the energy-substituted mass 
$m_{\rm ES} = \left [ (s/2+{\bf p}_i\cdot{\bf p}_B)^2/E_i^2-{\bf p}_B^2\right ] ^{1/2}$,
where $(E_i,{\bf p}_i)$ and $(E_B,{\bf p}_B)$ are the four-momenta
of the \FourS and $B$ candidate respectively, 
and the asterisk denotes the \FourS frame.
Our signal lies in the region $|\Delta E|\le 0.1$ GeV and 
$5.27 \le m_{\rm ES} \le 5.29$ GeV.  Sidebands in $m_{\rm ES}$ and $\Delta E$ 
are used to characterize the continuum background. 
The average number of signal $B$ candidates per selected data event ranges 
from 1.05 to 1.27, depending on the final state.
A single candidate per event is chosen 
as the one with the smallest $B$ vertex-fit $\chi^2$ 
(\rhopKzshort and \rhozKzshort), the smallest value of $\chi^2$ 
constructed from deviations of reconstructed $\pi^0$ masses
from the expected value (\rhopKmshort ), or randomly (\rhozKpshort ).
Monte Carlo (MC) simulation shows that up to 38\% (23\%) of longitudinally 
(transversely) polarized signal events are misreconstructed
with one or more tracks originating from the other $B$ in the event. 

To reject the dominant $q\overline q$ continuum background we require
$|\cos\theta_T|<0.8$, where $\theta_T$ is the c.m.\ frame angle between the
thrust axes of the $B$-candidate and that formed from the other tracks and
neutral clusters in the event. We also use as discriminant variables
the polar angles of the $B$-momentum vector and the $B$-candidate thrust
axis with respect to the beam axis, and the two Legendre moments $L_0$ and
$L_2$ of the energy flow around the $B$-candidate thrust axis
in the c.m.\ frame~\cite{bib:Legendre}.
These variables are combined in a Fisher discriminant ${\cal F}$ 
(\rhozKpshort ) or a neural network (NN) (other modes).
Finally, we suppress background from B decays to charmed states
by removing signal candidates that have decay products consistent with 
$D^0\rightarrow  K^-\pi^+ (\pi^0)$ and $D^-\rightarrow  K^+\pi^- \pi^-$
decays. 


We use an extended (not extended in the \rhopKzshort mode)
unbinned maximum-likelihood (ML) fit to extract signal yields,
asymmetries, and angular polarizations simultaneously. We define the likelihood
${\cal L}_i$ for each event candidate $i$ as the sum of 
$n_j {\cal P}_j(\vec x_i; \vec \alpha)$ over hypotheses $j$
(signal, $q\bar q$ background, and several \BB backgrounds discussed below),
where the ${\cal P}_j(\vec x_i; \vec \alpha)$ are the probability density 
functions (PDFs) for the measured variables $\vec x_i$, and $n_j$ are the
yields for the different hypotheses. 
The quantities $\vec \alpha$ represent parameters in
the expected distributions of the measured variables for each hypothesis.
They are extracted from MC simulation and
$(m_{\rm ES},\Delta E)$ sideband data. 
They are fixed in the fit except for some shape parameters of the continuum
$\Delta E$ and $m_{\rm ES}$ distributions.
The extended likelihood function for a sample of $N$ candidates is
${\cal L} = 
\exp{(-\sum n_j)}
\prod_{i=1}^{N} {\cal L}_i 
$.

The fit input variables $\vec x_i$ are $m_{\rm ES}$, $\Delta E$,
NN or ${\cal F}$, invariant masses of the candidates $\rho$ ($f_0 (980)$) and
$K^*$, and helicity angles $\theta_{\rho}$ and $\theta_{K^*}$.
We study large control samples of
$B\rightarrow D\pi$ decays of similar topology to verify the simulated 
resolutions in $\Delta E$ and $m_{\rm ES}$, adjusting the PDFs to account for
any difference found.

Since almost all correlations among the fit input variables are found
to be small, we take each ${\cal P}_j$ to be the product of the PDFs
for the separate variables with the following exceptions where we
explicitly account for correlations: the correlation between the two 
helicity angles in signal, the correlation due to misreconstructed events 
in signal, and the correlation between mass and helicity in backgrounds. 
The effect of neglecting other correlations is evaluated by fitting ensembles 
of simulated experiments in which we embed the expected numbers of signal and 
charmless $B$-background events, randomly extracted from fully-simulated MC 
samples.

\begin{figure}[htb]

\includegraphics[width=8.4cm]{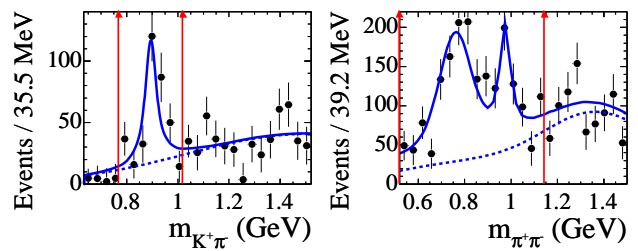}
\caption{sPlots for the invariant mass of
$K\pi$ in \rhopKzshort (left) and $\pi\pi$ in \rhozKzshort 
and \fKzshort (right) 
up to the higher-mass regions.
The points with error bars show the data, and
the solid (dashed) lines 
show the projected PDFs of the signal and non-resonant background 
(non-resonant background only: $\rho K\pi$ in \rhopKzshort ; the 
sum of $f_0(1370) {K^*}$, $\pi\pi K^*$, and $\pi\pi K\pi$ in \rhozKzshort ).
The arrows show the nominal fit regions.}
\label{fig:extended}
\end{figure}

We use MC-simulated events to study backgrounds from other $B$ decays. 
Charmless $B$-backgrounds are grouped into up to 11 classes with similar 
topologies depending on the mode.  
Yields for decays with poorly known branching fractions are 
varied in the fit with those remaining kept fixed to their measured values. 
One to four additional classes account for neutral and charged $B$ decays to 
final states with charm.
Up to 6 classes account for misreconstructed events in signal.
We also introduce components for non-resonant backgrounds such as 
$\pi\pi K^*$, $\rho K\pi$, $f_0(980) K\pi$, and $f_0(1370) K\pi$,
which differ from signal only in resonance mass and helicity 
distributions.
The magnitudes of these components are determined by extrapolating from
fits performed on a wider mass range reaching to higher mass values
and are fixed in the fit.
Fig.~\ref{fig:extended} shows the sPlots~\cite{bib:splot} 
for the invariant mass of $K\pi$ and $\pi\pi$ 
in the \rhopKzshort and \rhozKzshort modes, respectively.
The data events are weighted by their probability to be signal,
calculated from the signal and backgrounds PDFs of the 
\DeltaE, \mes, and NN variables.

\begin{figure}[htb]
\includegraphics[width=9.0cm]{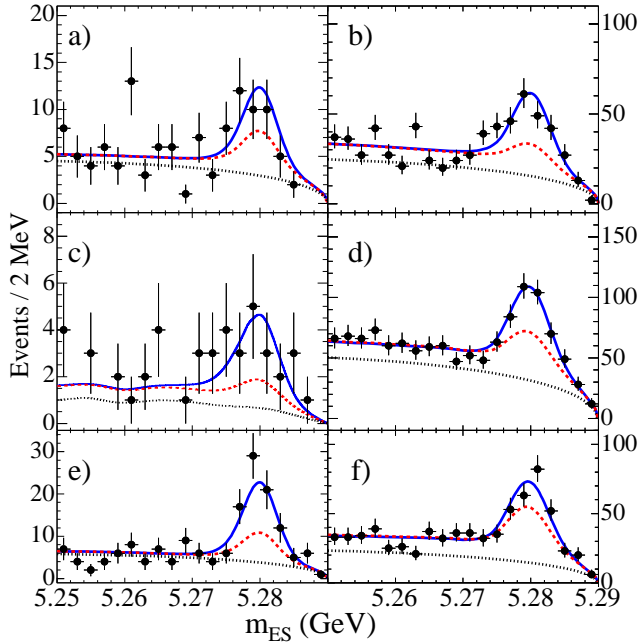} 
\caption{Projections of the multidimensional fit onto
$m_{\rm ES}$ for events passing a 
signal-to-total likelihood probability ratio cut
with the plotted variable excluded for
(a) \rhozKpshort ,
(b) \rhopKzshort ,
(c) \rhopKmshort ,
(d) \rhozKzshort , 
(e) \fKpshort , and
(f) \fKzshort .
The points with error bars show the data;
the solid, dashed and dotted lines show the total,
background, and continuum PDF projections
respectively.}
\label{fig:projections}
\end{figure}

The results of the ML fits are summarized in 
Table~\ref{tab:results}. For the branching fractions, we assume
equal production rates of \BpBm and \BzBzb. 
The significance $S$ of a signal is defined by $\Delta\ln {\cal L} = S^2/2$, 
where $\Delta\ln {\cal L}$ represents the change in likelihood from the maximal 
value when the number of signal events is set to zero,
corrected for the systematic error defined below.
We find significant signals for \rhopKzshort , \rhozKzshort , and
\fKpshort, and some evidence for \fKzshort . For the modes with significance
smaller than five standard deviations we also measure the 90\%
confidence level (C.L.) upper limit, taking into account the systematic
uncertainty.
Fig.~\ref{fig:projections} shows projections of the fits onto $m_{\rm ES}$.

\begin{table*}[htb]
\caption{Summary of results for the measured $B$-decay modes:
signal yield $n_{sig}$ and its statistical uncertainty,
reconstruction efficiency  $\varepsilon$,
daughter branching fraction product $\prod {\cal B}_i$,
significance S (systematic uncertainties included),
measured branching fraction ${\cal B}$,
(90\% C.L. upper limit in parentheses),
measured longitudinal polarization $f_L$ 
(for the modes with non-significant signals
the numbers, in brackets, are not quoted as measurements)
and charge asymmetry ${\cal A}_{\rm CP}$.}
\begin{center}
\begin{tabular}{lcccccccc}
\hline
\hline
Mode  &
$n_{sig}$   &  
$\varepsilon$(\%)  & 
$\prod {\cal B}_i$(\%) & 
$S (\sigma)$ &
${\cal B}(10^{-6})$  & 
$f_L$  &
${\cal A}_{\rm CP}$ \\
\hline
\hline
\rhozKpshort                               &                     &       &        &  2.5  &  $3.6^{+1.7}_{-1.6}\pm 0.8$  (6.1) & [$0.9\pm 0.2$]        &  -- \\ 
\hspace*{0.3cm}$\rightarrow$\rhozKpshorta  &  $19^{+16}_{-15}$   &  7.9  &  32.9  &  1.3  &  $3.2^{+2.7}_{-2.4}\pm 0.9$        & [$0.8^{+0.3}_{-0.5}$] &  -- \\ 
\hspace*{0.3cm}$\rightarrow$\rhozKpshortb  &  $32^{+19}_{-17}$   & 15.8  &  22.9  &  2.1  &  $3.8^{+2.2}_{-2.1}\pm 0.9$        & [$1.0\pm{0.3}$]       &  -- \\
\hline
\rhopKzshort   &  $194\pm 29$  &  13.5  &  66.7  &  7.1  &  $9.6\pm 1.7\pm 1.5$  & $0.52\pm 0.10\pm 0.04$  &  $-0.01\pm 0.16 \pm 0.02$ \\  
\hline
$\rho ^- {K^*}^+_{K^+\pi^0}$ &  $60^{+25}_{-22}$  &  15.2  &  32.5  &  1.6  &  $5.4^{+3.8}_{-3.4}\pm 1.6$ (12.0)  &  [$-0.18^{+0.52}_{-1.74}$] &  -- \\  
\hline
\rhozKzshort   &  $185\pm 30$  &  22.9  &  66.7  &  5.3  &  $5.6\pm 0.9 \pm 1.3$ & $0.57\pm 0.09 \pm 0.08$ &  $0.09\pm 0.19 \pm 0.02$ \\  
\hline
\fKpshort                                  &                     &       &        &  5.0  &  $5.2\pm 1.2\pm 0.5$               &  --  & $-0.34\pm 0.21 \pm 0.03$  \\ 
\hspace*{0.3cm}$\rightarrow$\fKpshorta     &   $40^{+13}_{-12}$  &  8.5  &  32.9  &  3.8  &  $6.2^{+2.1}_{-1.9}\pm 0.7$        &  --  & $-0.50\pm 0.29 \pm 0.03$          \\  
\hspace*{0.3cm}$\rightarrow$\fKpshortb     &   $37^{+14}_{-12}$  & 16.6  &  22.9  &  3.2  &  $4.2^{+1.5}_{-1.4}\pm 0.5$        &  --  & $-0.13\pm 0.30 \pm 0.01$  \\  
\hline
\fKzshort      &  $83\pm 19$   &  21.7  &  66.7  &  3.5  &  $2.6\pm 0.6 \pm 0.9$ (4.3) &  -- &  $-0.17\pm 0.28 \pm 0.02$ \\  
\hline
\hline
\end{tabular}
\label{tab:results}
\end{center}
\end{table*}

A source of systematic error is related to the determination of the PDFs 
and is due to the limited statistics of the Monte-Carlo and to 
the uncertainty on the PDF shapes. We obtain variations in the yields 
ranging from 1 to 18\%, depending on the mode.
The systematic error due to the non-resonant background extrapolation 
and interference with signal is in the range 6--21\%. 
Event yields for $B$-background modes fixed in the fit are
varied by their respective uncertainties. This results in a systematic
uncertainty of 2--12\%.
We evaluate and correct for possible fit biases with MC experiments.
We assign a systematic uncertainty of  1--7\% for this.

\begin{figure}[htb]
\begin{center}
  \includegraphics[width=8.7cm]{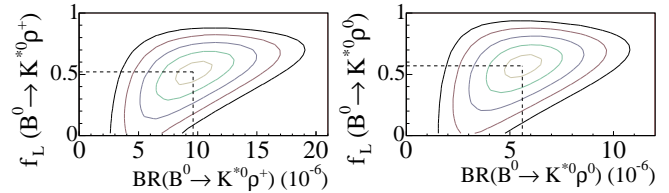} 
\end{center}
\caption{Distribution of $-2\ln {\cal L} ({\cal B},f_L)$ for 
\rhopKz (left) and \rhozKz (right) decays. The solid dots correspond
to the central values and the curves give contours in
$\Delta \sqrt{-2\ln {\cal L} ({\cal B},f_L)}=1$ steps.}
\label{fig:contour}
\end{figure}

The reconstruction efficiency depends on the decay polarization.
For the \rhozKpshort mode we calculate the efficiency using the 
measured polarization (combined for the two \rhozKpshort modes) 
and assign a systematic uncertainty corresponding to the total
polarization measurement error (9 and 20\% for each mode respectively).
For the other modes we exploit the correlation
between ${\cal B}$ and $f_L$ 
and obtain the values of ${\cal B}$ from fits where 
${\cal B}$ and $f_L$ are free parameters. 
Fig.~\ref{fig:contour} shows the behavior of $-2\ln {\cal L}(f_L, {\cal B})$
for the modes with significant signal.

Additional reconstruction efficiency uncertainties arise from tracking
(3--5\%), particle identification (1--2\%), 
vertex probability (2\%), track multiplicity (1\%) and thrust angle (1\%).
$K^0_S$ and $\pi^0$ reconstruction contribute 2.3\% and 3\% uncertainty, 
respectively. Other minor systematic effects are from uncertainty in 
daughter branching fractions, MC samples statistics, and number of $B$ mesons.
The absolute systematic uncertainty in $f_L$ takes 
into account PDF shape variations (5--10\%), 
$B$ and non-resonant backgrounds (4--8\%),
and efficiency dependence on the polarization (1--2\%). 
The absolute uncertainty in the charge asymmetry due to track charge bias
is less than 1\%. PDF variations and fixed $B$-background effects 
contribute up to 2\%.

In summary, we have searched for 
$B \rightarrow \rho K^*$ and $B \rightarrow f_0(980)K^*$\ decays.
We observe \rhopKz , \rhozKz , \fKp , and \fKz\ with 7.1, 5.3, 5.0, and 3.5 
$\sigma$ significance respectively. 
We measure the branching fractions or 90\% C.L. upper limits, 
the fractions of longitudinal polarization, and the charge asymmetries,
summarized in Table~\ref{tab:results}.
The measured polarization in the \rhopKzshort\ and \rhozKzshort\ 
modes agrees with values measured in $\phi K^*$ decays.

We thank I. Bigi, S. Descotes-Genon, O. P\`{e}ne, and M. Pennington for their
advice on the treatment of non-resonant backgrounds.
We are grateful for the excellent luminosity and machine conditions
provided by our \pep2\ colleagues, 
and for the substantial dedicated effort from
the computing organizations that support \babar.
The collaborating institutions wish to thank 
SLAC for its support and kind hospitality. 
This work is supported by
DOE
and NSF (USA),
NSERC (Canada),
IHEP (China),
CEA and
CNRS-IN2P3
(France),
BMBF and DFG
(Germany),
INFN (Italy),
FOM (The Netherlands),
NFR (Norway),
MIST (Russia),
MEC (Spain), and
PPARC (United Kingdom). 
Individuals have received support from the
Marie Curie EIF (European Union) and
the A.~P.~Sloan Foundation.


\bibliographystyle{h-physrev2-original}  

\end{document}